\documentclass[
 aps,
 11pt,
 final,
 notitlepage,
 oneside,
 twocolumn,
 nobibnotes,
 nofootinbib,
 superscriptaddress,
 noshowpacs,
 centertags]
{revtex4}

\def\saoname{Special Astrophysical Observatory,  Russian Academy of Sciences,
              Nizhnii Arkhyz, 369167 Russia}

\def\squareforqed{\hbox{\rlap{$\sqcap$}$\sqcup$}}

\def\sq{\ifmmode\squareforqed\else{\unskip\nobreak\hfil
\penalty50\hskip1em\null\nobreak\hfil\squareforqed
\parfillskip=0pt\finalhyphendemerits=0\endgraf}\fi}

\def\degr{\hbox{$^\circ$}}

\def\arcsec{\hbox{$^{\prime\prime}$}}

\def\utw{\smash{\rlap{\lower5pt\hbox{$\sim$}}}}

\def\udtw{\smash{\rlap{\lower6pt\hbox{$\approx$}}}}

\def\diameter{{\ifmmode\mathchoice
{\ooalign{\hfil\hbox{$\displaystyle/$}\hfil\crcr
{\hbox{$\displaystyle\mathchar"20D$}}}}
{\ooalign{\hfil\hbox{$\textstyle/$}\hfil\crcr
{\hbox{$\textstyle\mathchar"20D$}}}}
{\ooalign{\hfil\hbox{$\scriptstyle/$}\hfil\crcr
{\hbox{$\scriptstyle\mathchar"20D$}}}}
{\ooalign{\hfil\hbox{$\scriptscriptstyle/$}\hfil\crcr
{\hbox{$\scriptscriptstyle\mathchar"20D$}}}}
\else{\ooalign{\hfil/\hfil\crcr\mathhexbox20D}}%
\fi}}

\newcommand{\aap}{Astron. and Astrophys. }
\newcommand{\aaps}{Astron. and Astrophys. Suppl. }
\newcommand{\aj}{Astron.~J. }
\newcommand{\apjs}{Astrophys.~J. Suppl. }
\newcommand{\bsao}{Bull. Spec. Astrophys. Obs. }
\newcommand{\mnras}{Monthly Notices Royal Astron. Soc. }
\newcommand{\pasp}{Publ. Astron. Soc. Pacific }
\newcommand{\alet}{Astronomy Letters }
\newcommand{\an}{Astronomische Nachrichten }
\begin{document}
%\selectlanguage{english}

%\flushbottom \columntovsizetrue
\allowdisplaybreaks\selectlanguage{english} 
%\input engnames

%\keywords{astronomical data bases---galaxies: general}

%\titlerunning{DATABASE FOR STUDYING EDGE-ON GALAXIES}

%\authorrunning{MAKAROV, ANTIPOVA}

%\toctitle{Database For Studying Edge-on Galaxies} \tocauthor{}

\title{Database For Studying Edge-on Galaxies}

\author{\firstname{D.~I. }~\surname{Makarov}} \email{dim@sao.ru} \affiliation{\saoname}

\author{\firstname{A.~V.}~\surname{Antipova}} \affiliation{\saoname}

\received{August 3, 2020 } \revised{December 10, 2020} \accepted{December 10, 2020}

\begin{abstract}
We present a database created within the project on studying
edge-on galaxies. These galaxies provide a unique opportunity to
study the three-dimensional distribution of the matter in galaxy
disks, which is extremely important for analyzing the influence of
internal and external factors on the evolution of galaxies. For
the moment, extensive observed material has been accumulated on
the kinematics and photometry of such galaxies. The database is
designed to organize information, make it easier to visualize, and
to improve works on studying this type of objects. The database
combines information from previous catalogs on edge-on galaxies
and data from current projects; provides access to astrometric and
photometric data; carries out interconnection with other
databases. The present paper describes the structure and
web-access to the database: \url{https://www.sao.ru/edgeon/}.
\end{abstract}

\maketitle

\section{INTRODUCTION}

Galaxies seen at an angle close to $90\degr$ provide a unique
opportunity to study the vertical structure of galaxy disks and
bulges. Thus, they traditionally attract the attention of
researchers. P.~van~der~Kruit and K.~Freeman with colleagues
performed a number of fundamental studies on such galaxies in
1980--1990. They had shown that the vertical distribution of the
matter in galaxy disks is well described by the self-gravitating
isothermal layer model, in which the vertical disk scale is
directly related to the vertical velocity dispersion and surface
density of stars in the disk \citep{van_der_Kruit1981}. It was
found that the vertical scale of the galaxy disks remains
practically constant along the radius of the galaxy.

The description of the dust influence in edge-on galaxies is a
difficult task. It is necessary to take into account the complex
structure of a galaxy, differences in the radial and vertical
distribution of stars and dust, radiation scattering on dust
particles. \citet{Dalcanton2004} found that the dust lane only
occurs in rapidly rotating systems, $V_{\rm rot}>120$~km s$^{-1}$.
Slowly rotating galaxies show a diffuse dust distribution. The
thickness of layers of the dust and gas in dwarf, \mbox {$V_{\rm
rot}<100$~km s$^{-1}$}, thin galaxies is comparable to the
vertical scale of the stellar disk \citep{Bizyaev2017}. The
occurrence of the dust lane in massive galaxies is most likely
associated with gravitational instability in the mixed
gaseous-stellar disk and can lead to a significant increase of the
star-formation rate in massive galaxies.

Edge-on spirals of late morphological types are flat galaxies with
the axis ratio $a/b>7$. As a rule, such galaxies are gas-rich,
have a low surface brightness and a low star-formation rate
\citep{Matthews1999, Matthews2000, Kautsch2009}.
\citet{Kudrya1994}, based on the distribution of apparent
ellipticity of flat galaxies, showed that the maximum true axial
ratio in the blue range for galactic disks is $a/b=25.8$. In the
papers by \citet{Zasov1991}, \citet{Sotnikova2006}, and
\citet{Khoperskov2010}, it is emphasized that the existence of
very thin, $a/b>10$, purely disk galaxies is possible only in the
presence of a massive dark halo around them. The theoretical ratio
between the thickness of the disk and the mass of its spherical
component allows one to estimate a lower boundary of the dark halo
mass in galaxies. An analysis of the rotation curves of ultra-thin
galaxies indicates the existence of the dark matter halo with a
compact core, where the characteristic radius of the dark halo
core is approximately equal to half the scale of the galaxy disk
\citep{Kurapati2018}.

The above results urge the importance of studying edge-on
galaxies, of understanding the physical processes of formation,
evolution, and morphological transformation of galaxies in the
Universe.

Databases have proved themselves as an important tool for
collecting, organizing, and analyzing various information. To the
date, the most famous and largest are: the NASA Extragalactic
Database\footnote{\url{https://ned.ipac.caltech.edu/}} (NED),
HyperLeda\footnote{\url{http://leda.univ-lyon1.fr/}}
\citep{HyperLeda},
SIMBAD\footnote{\url{http://simbad.u-strasbg.fr/simbad/}}, and
Vizier\footnote{\url{https://vizier.u-strasbg.fr/viz-bin/VizieR}}.
These databases contain data from various sky surveys and
observations of individual objects published in the literature;
they provide an interface for searching for, identifying, and
visualizing a variety of information; they interconnect different
databases. However, the versatility and inability to ``boil the
ocean'' lead to certain difficulties and limitations of these
resources when studying specific types of objects like a sample of
edge-on galaxies. In particular, automatic algorithms for the
selection, classification, and photometry of such objects often
fail. The distribution of galaxies by size, morphology, and
surface brightness is extremely wide. This results in the
impossibility of choosing the optimum set of parameters that would
equally well describe both extremely distant, practically point
objects at cosmological redshifts and nearby extended galaxies
showing a complex and rich internal structure. As a result, there
are well-known problems of artificial splitting of extended
objects into smaller ones and, as a consequence, incorrect
estimation of their observation parameters. Edge-on galaxies
occupy a specific place in this row, since the effects of
integration along the line of sight, the presence of an extremely
strong dust lane for giant galaxies, and the considerable
ellipticity of objects play an important role in the distribution
of their surface brightness. All these facts lead to large
systematic errors of the observed parameters of edge-on galaxies
stored in universal databases.

Recently, edge-on galaxies have been actively studied including
by the authors of the paper. The creation of this database was
caused by the personal requirements of the authors in structuring
and systematization of data, in the convenience of their
visualization, and in facilitating the analysis of the results.
The need arose to have at hand a universal tool for working with
samples of galaxies, which could be easily and quickly adapted to
various tasks solved when studying edge-on galaxies. This
determined the motivation of creating a database of objects of
this type.

\section{DESCRIPTION OF THE DATABASE}

\begin{figure*}
\setcaptionmargin{5mm} \onelinecaptionstrue \captionstyle{normal}
\includegraphics[width=\textwidth]{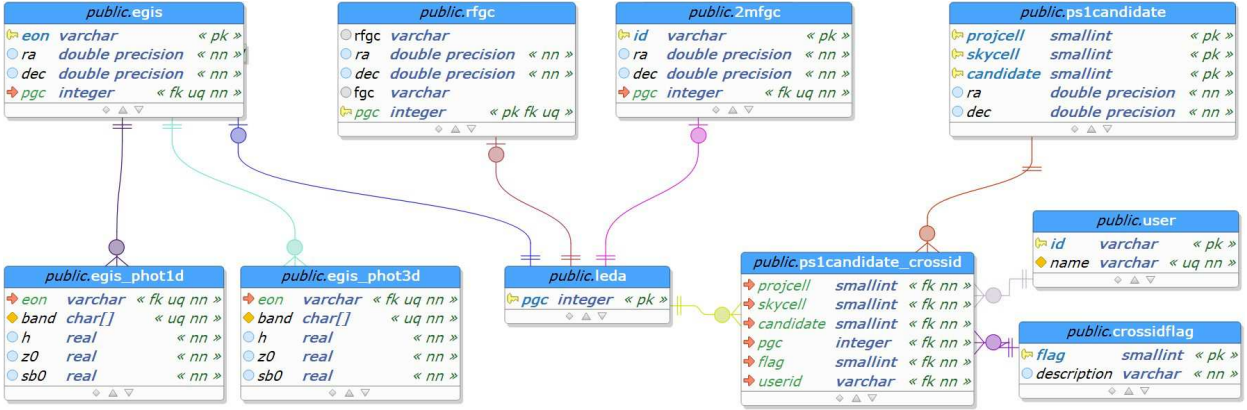}
\caption{Database structure schematic diagram. The structure of
Tables is described in more detail in the Appendices.}
\label{fig:DBSchema}
\end{figure*}

The database for studying edge-on galaxies operates on the servers
of the Special Astrophysical Observatory of the Russian Academy of
Sciences: \url{https://www.sao.ru/edgeon/}.

The database operates under the control of the PostgreSQL
object-relational database management
system\footnote{\url{https://www.postgresql.org/}}. This open
source system is characterized by reliability, functionality, and
performance. PostgreSQL is based on the structured query language
SQL. PostgreSQL provides support for embedded procedural
programming languages: PL/pgSQL, PL/Perl, PL/Python, and PL/Tcl.
It is possible to load extension modules in C. PostgreSQL provides
APIs for a wide variety of programming languages such as Python,
Perl, PHP, ODBC, etc. It is characterized by ease of expansion,
the ability to create new data types, indices, operators, and
functions. PostgreSQL also implements table inheritance.

The database is based on three catalogs of edge-on galaxies. These
are: the Revised Flat Galaxy Catalogue (RFGC) \citep{RFGC}, the
Catalog of Edge-on Disk Galaxies from SDSS (EGIS), and the
2MASS-selected Flat Galaxy Catalog (2MFGC). The basic structure of
the database is shown in Fig. \ref{fig:DBSchema} illustrating the
relations between various tables.

\subsection{Revised Flat Galaxy Catalogue}

The RFGC catalog was created in 1999 by Igor~D.~Karachentsev and
his colleagues and contains data on 4236 thin galaxies with
diameters $a>40\arcsec$ and the axial ratio $a/b\ge7$ in ``blue''
photographic images of POSS and ESO/SERC sky surveys. In addition,
with the purpose of continuity, it included 208 objects from
previous versions of the FGC \citep{FGC} catalog, which ceased to
meet the selection criterion after the refinement of their
parameters.

The catalog contains the data on the positional angle of a galaxy;
the ``blue'' and ``red'' sizes obtained from the POSS-I prints;
the total apparent $B$ magnitude calculated with the dimensions,
morphological type, and surface brightness type, as described in
the RFGC \citep{RFGC} catalog; the morphological type of a galaxy;
the asymmetry index; the surface brightness index and the number
of significant satellites. The detailed structure of the catalog
is given in Appendix.~\ref{a:RFGC}.

\subsection{Catalog of Edge-on Disk Galaxies from SDSS}

As the name suggests, the EGIS catalog was created based on the
SDSS \citep{EGIS} survey and contains the data on 5747 true
edge-on galaxies. The catalog consists of three tables:
\begin{list}{}{
\setlength\leftmargin{4mm} \setlength\topsep{2mm}
\setlength\parsep{0mm} \setlength\itemsep{2mm} } \item {\it
egis}---the list of galaxies with their identification and
astrometry; \item {\it egis\_phot1d}---the photometric parameters
of galaxies obtained from the analysis of the one-dimensional
profile of a galaxy. The table gives the horizontal exponential
scale of the disk; the vertical $\mathrm{sech}^2$ scale of the
disk; the central surface brightness normalized to the face-on
position of a galaxy; the total aperture magnitude corrected for
extinction in our Galaxy, and the contribution of the bulge to the
total luminosity of a galaxy; \item {\it egis\_phot3d}---the disk
parameters (the central surface brightness, the vertical and
horizontal scales) obtained from modeling the surface brightness
distribution in the galaxy image. \end{list} Photometric
parameters were derived from the images obtained in three filters,
$gri$, from the SDSS\,dr7 survey \citep{Abazajian2009}. The
catalog is provided by an archive of the processed images in the
corresponding filters used in the photometry. The structure of the
catalog is described in Appendix~\ref{a:EGIS}.

\subsection{2MASS-Selected Flat Galaxy Catalog}

The 2MFGC catalog was created based on the automatic selection of
objects from the 2MASS infrared sky survey \citep{2MFGC}. It
contains 18\,020 galaxies distributed over the whole sky with the
axial ratio  $a/b\ge3$. The catalog contains information on
photometry in the $J$, $H$, and $K_s$ bands from the Extended
Source Catalog of the 2MASS survey \citep{Jarrett2000}. The
structure of the catalog is described in Appendix~\ref{a:2MFGC}.

\subsection{Edge-on Galaxy Candidates from the Pan-STARRS Survey} \label{s:PS1candidate}

\begin{figure*}
\setcaptionmargin{5mm} \onelinecaptionstrue \captionstyle{normal}
\includegraphics[width=\textwidth]{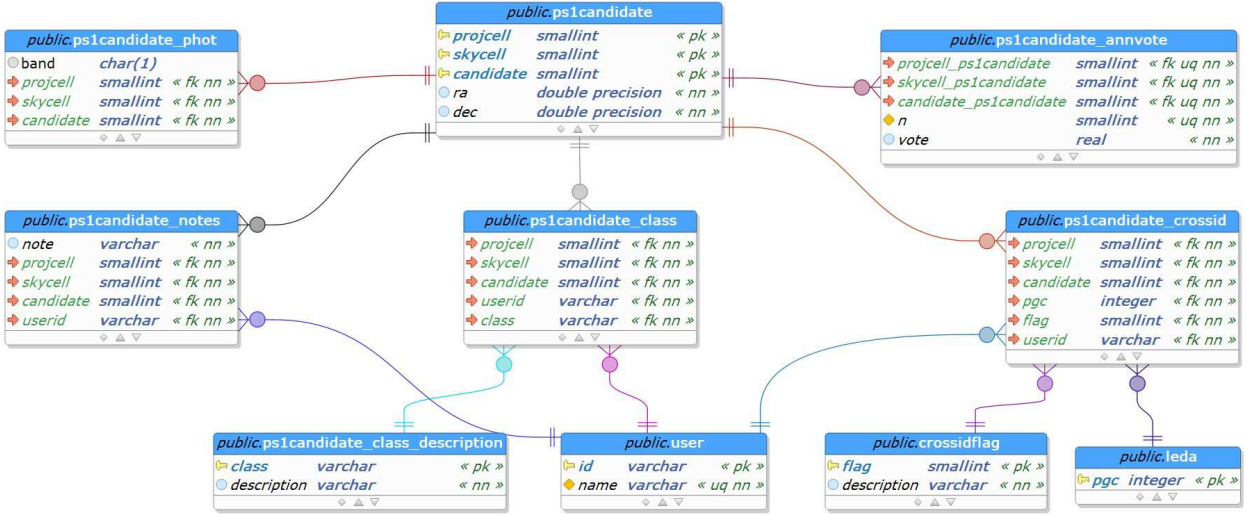}
\caption{Schematic diagram of the structure of the catalog of
edge-on galaxy candidates from Pan-STARRS. The structure of tables
is described in more detail in Appendix~\ref{a:Pan-STARRS}.}
\label{fig:PS1Schema}
\end{figure*}

Our database is actively used in the work to create a new
catalog of edge-on galaxies based on the Pan-STARRS sky survey. We
are preparing this catalog and the search algorithm for galaxies
included in it for publication. Here we provide a brief
description and data structure of the candidate catalog. All
information about the current state of the project is available on
the project
web-page\footnote{\url{https://www.sao.ru/edgeon/catalogs.php?cat=PS1candidate}}.

Approximately 27\,000 candidates were selected using artificial
neural networks (ANNs) trained on a sample of galaxies from EGIS
and RFGC catalogs. Subsequently, all candidates visually
classified to eliminate false objects and various artefacts
disguised as edge-on galaxies. At this stage, we are refining the
photometry of the selected candidates. The database was used to
store and organize the data on candidates; to provide a convenient
interface for viewing the data of specific objects; to screen out
the image defects disguised as the target galaxies; to visually
classify the objects (see Section~\ref{s:VisualClassification});
to compile various samples and analyze the data.

Due to the complexity of the project and its multistage structure,
the catalog structure is more ramified than the above-described
already-present catalogs. The schematic diagram of the catalog in
the database is illustrated in Fig. \ref{fig:PS1Schema}. For the
sake of compactness, it only lists the panels important for
understanding the relationship between tables. The catalog of
candidates consists of the following tables:
\begin{list}{}{
\setlength\leftmargin{4mm} \setlength\topsep{2mm}
\setlength\parsep{0mm} \setlength\itemsep{2mm} } \item {\it
ps1candidate}---the list of candidates; \item {\it
ps1candidate\_annvote}---the classification performed using ANNs;
\item {\it ps1candidate\_class}---the results of visual inspection
of candidates; \item {\it ps1candidate\_phot}---the automatic
photometry performed by
SExtractor\footnote{\url{https://sextractor.readthedocs.io/}};
\item {\it ps1candidate\_crossid}---the cross-identification with
the galaxies from the HyperLeda database \citep{HyperLeda}; \item
{\it ps1candidate\_notes}---the various notes made while working
with galaxies.
\end{list}

To facilitate the work with the candidate catalog, views were
developed that ``on-the-fly'' combine the data from various tables
into a single structure:
\begin{list}{}{
\setlength\leftmargin{4mm} \setlength\topsep{2mm}
\setlength\parsep{0mm} \setlength\itemsep{2mm} } \item {\it
ps1candidate\_class\_stats} calculates statistics on the visual
classification of each galaxy; \item {\it ps1candidate\_final} is
the ``cleared'' sample of objects to create the final version of
the catalog.
\end{list}

The structure of the catalog is described in
Appendix~\ref{a:Pan-STARRS}.

The catalog is accompained by an archive of the images. At the
moment, it contains  \mbox {FITS files} in five filters taken from
the archive of
Pan-STARRS\footnote{\url{https://ps1images.stsci.edu/cgi-bin/ps1cutouts}}
and JPEG images with the outlines of the selected objects. \mbox
{JPEG images} are used for data visualization using the web
interface.

\subsection{Interaction with HyperLeda}

The interrelation between objects in different catalogs within our
database is carried out using a unique identifier: the PGC numbers
of galaxies in the HyperLeda
database\footnote{\url{http://leda.univ-lyon1.fr/}}
\citep{HyperLeda}. This allows one to link together the
galaxy-specific data from various tables and catalogs and
visualize the whole dataset. We carry out cross-correlation of
objects with galaxies from the HyperLeda database
\citep{HyperLeda}. In the absence of matches, the missing galaxies
are added to the HyperLeda database, which allows one, on the one
hand, to keep the interrelations up to date and, on the other
hand, to replenish HyperLeda with new data.

\section{WEB INTERFACE}

The home page\footnote{\url{https://www.sao.ru/edgeon/}} provides
information about the project, describes its goals and objectives.
The ``Projects'' and ``Catalogs'' Sections are dedicated to
current projects and data collection. The data-visualization web
interface is implemented in the form of two universal options:
visualization of the catalog as a whole in the form of a table and
visualization of information cards of individual objects with the
detailed data. In addition to universal visualization methods,
each specific dataset can have its own display system for more
flexible adaptation to the specific features of the structure of
the catalogs and the data they contain. The search for objects in
the database by coordinates is also provided. The web interface of
our edge-on galaxy database is implemented in the
PHP\footnote{\url{https://www.php.net/}} scripting language using
JavaScript. The system is constantly expanding and supplemented
with new data and capabilities.

\subsection{Service of Visual Classification of Objects} \label{s:VisualClassification}

\begin{figure*}
\setcaptionmargin{5mm} \onelinecaptionstrue \captionstyle{normal}
\includegraphics[width=\textwidth]{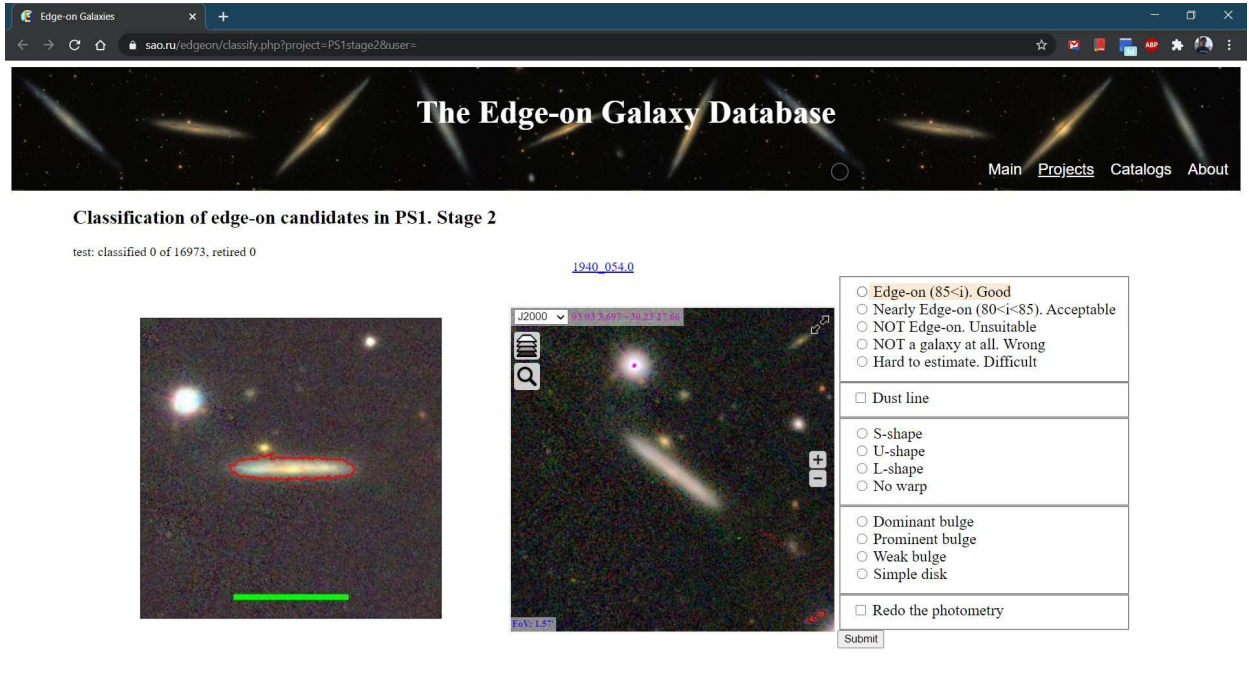}
\caption{Example of the web page with the service of visual
classification of galaxies. Each user is informed how many objects
from the list he has classified and for how many objects
sufficient statistics have been collected (retired). A web link to
the page of the object itself is given (1940\_054.0). An image of
a galaxy with the outline (the left-hand picture) and a
corresponding region of the sky map visualized using the Aladin
Lite code (the right-hand picture) are given. On the right, is a
classification section that allows one to group questions by
topic.} \label{fig:VisualClassification}
\end{figure*}

Working with the catalog of edge-on galaxy candidates (see
Section~\ref{s:PS1candidate}) demanded the large-scale visual
inspection of images, object classification, elimination of
``waste'', and storing information in the database. Initially, for
this purpose, we used the
Zooniverse\footnote{\url{https://www.zooniverse.org/}} citizen
science service  developed from the Galaxy Zoo project. However,
this appeared labor intensive in preparing images for viewing,
uploading them to the Zooniverse servers, organizing the
classification itself, and then analyzing the results. Therefore,
we have developed our own visual inspection and classification
system for galaxies deployed on our server. This greatly
simplified and accelerated the process of creating a
questionnaire, conducting a visual inspection, and analyzing the
results obtained. Our classification system is ideologically close
to the inquiry system in the Zooniverse project. The main
difference is associated with the active use of the Aladin
Lite\footnote{\url{https://aladin.u-strasbg.fr/}} code which
allows one to visualize this region of the sky from different
views, to scale images for a more detailed examination, and also
to freely move around the whole sky to control the surroundings of
the object.

The system allows one to quickly and easily create questionnaires,
link them to different datasets, and collect responses from the
registered users. At the moment, the visual classification service
does not imply the public access. Therefore, the responsibility
for forming a sample of objects for classification, creating a
system of questions, and providing access to project participants
rests with the system administrator. For this purpose, the
following tables have been created:
\begin{list}{}{
\setlength\leftmargin{4mm} \setlength\topsep{2mm}
\setlength\parsep{0mm} \setlength\itemsep{2mm} } \item {\it
user}---the list of users to personalize the work performed; \item
{\it quiz} contains a description of a specific project, links it
to tables in the database, and sets the limit on the number of
classifications per object; \item {\it quiz\_question}---the list
of questions that form the basis of the classification; \item {\it
quiz\_result}---the collection of classifications made by
different users in different projects; \item {\it
quiz\_result\_info}---the view combining classification
information ``on-the-fly'' and facilitating information analysis.
\end{list}

The scheme of tables is given in
Appendix~\ref{a:VisualClassification}. The web interface forms a
questionnaire based on information about the current project.
Questions can be grouped for easy classification. It is possible
to either select one item from a set of questions using the radio
buttons, or indicate the presence or absence of a specific feature
using flags. Information about the selected options is transferred
to the database, where it is written to Table \verb|quiz_result|
with an indication of the specific project, the user who performed
the classification, and the time. Objects for classification are
selected randomly from the list of those that have not yet been
classified by this user and the total number of classifications of
which has not exceeded the specified limit. An example of a
classification page is shown in Fig.
\ref{fig:VisualClassification}. This system was used in the visual
inspection of edge-on galaxy candidates in the Pan-STARRS survey.

\subsection{Service of Object Identification with the HyperLeda Database} \label{s:Identification}

\begin{figure*}
\setcaptionmargin{5mm} \onelinecaptionstrue \captionstyle{normal}
\includegraphics[width=\textwidth]{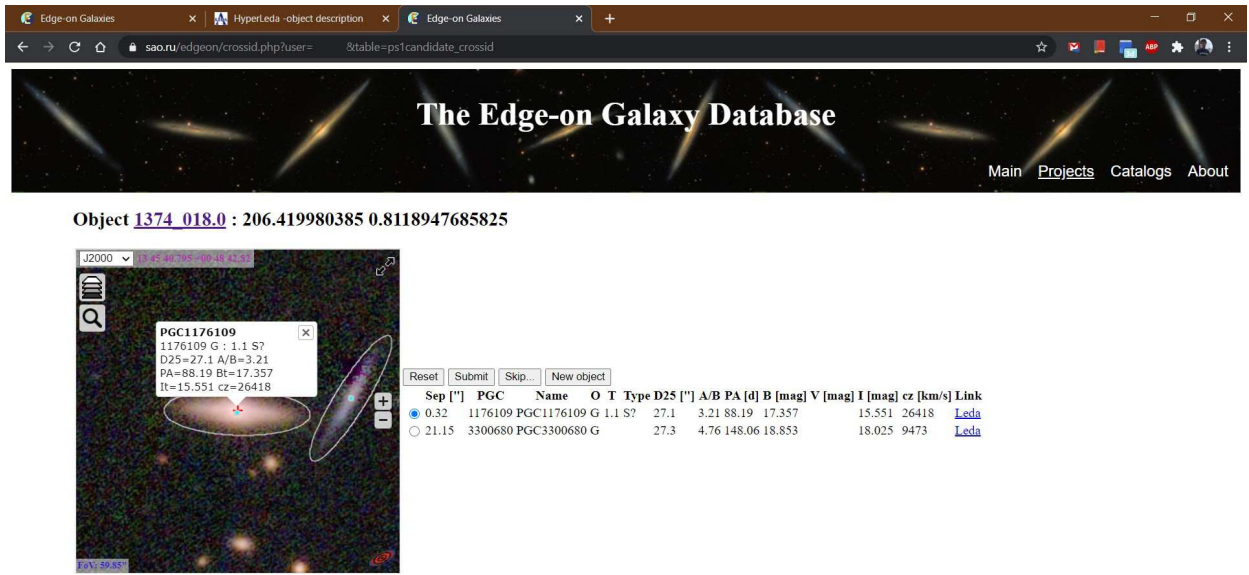}
\caption{Example of the web page with the service of visual
identification of objects. A web link is given to the object page
(1374\_018.0) and its coordinates in degrees. The Aladin Lite code
shows the neighborhood of an object in the sky.}
\label{fig:Identification}
\end{figure*}

The database implements a simple object identification
algorithm,similar to that adopted in HyperLeda \citep{HyperLeda}.
Two circles with the radii $R_1 < R_2$ are drawn around each
identified object. If only one HyperLeda object is found inside
these circles and it lies inside the $R_1$ circle, then this
object is automatically identified with the corresponding galaxy.
If no galaxy is found inside the $R_2$ circle in HyperLeda, then
the identified object is marked as a new galaxy. In all other
cases, the object is passed for manual verification. For this, the
web interface was organized, the work with which is regulated by
the administrator. An example of an identification page is shown
in Fig. \ref{fig:Identification}. Registered users are given the
opportunity to choose the best match between the identified object
and the list of known galaxies in HyperLeda presented in a tabular
form with the parameters most useful in the identification. For
convenience, well-known galaxies are sorted with distance from the
identified object. Aladin Lite is used to visualize the
surroundings of the object. The image shows both the position of
the identified object (marked with a red cross) and known galaxies
(blue dots), the characteristic sizes of which are enclosed in
white ellipses. The user can identify the object, or mark it as a
new galaxy, or postpone for further more detailed consideration.
The corresponding selection is entered into the database with
information about the user and the time of identification.

\section{CONCLUSION}

We have created a database for studying edge-on galaxies,
\url{https://www.sao.ru/edgeon/}. It is based on well-known
catalogs: RFGC \citep{RFGC}, EGIS \citep{EGIS}, and 2MFGC
\citep{2MFGC}. The system is easy to modify. This system was the
basis for a project to create a new catalog of edge-on galaxies
selected in the Pan-STARRS survey using artificial neural
networks. The database provides a web interface for accessing
catalogs and various data on galaxies such as astrometry,
photometric parameters, data on morphology, and visual
classification of objects. The interaction with the HyperLeda
\citep{HyperLeda} database and digital data archives through
Aladin Lite has been carried out.

This work is the first step in collecting and organizing data on
edge-on galaxies. Our nearest plans are to add a catalog of
photometric parameters obtained during a large-scale image
analysis using the
GalFit\footnote{\url{https://users.obs.carnegiescience.edu/peng/work/galfit/galfit.html}}
\citep{Peng2010} code from various sky surveys such as SDSS,
Pan-STARRS, and Legacy Survey. At the moment, the vast observed
material has been accumulated on the kinematics of gas in flat
galaxies, including observations that were carried out at the 6-m
SAO RAS telescope. The integration of this information into our
database is extremely important. Among the catalogs, with which we
intend to replenish our system, we should note a sample of edge-on
galaxies found in the deep fields of the Hubble telescope
\citep{Reshetnikov2019}. Comparison of the properties of galaxies
in the Local Universe and the galaxies at $z\sim1$ is extremely
important for understanding the evolution and morphological
transformation of galaxies. In addition, we plan to add a variety
of the data published in the literature.

We are sure that the presented database for the study of edge-on
galaxies will contribute to obtaining new astrophysical results.

\begin{acknowledgments}
We confirm that we used the HyperLeda
database\footnote{\url{http://leda.univ-lyon1.fr}}
\citep{HyperLeda}. The project used the Aladin Sky
Atlas\footnote{\url{https://aladin.u-strasbg.fr/}} developed at
CDS \citep{Aladin, AladinLite}.
\end{acknowledgments}

%\bibliographystyle{aspb1}
%\bibliography{Paper}
\section*{FUNDING}
The study was conducted with the financial support of the RFBR within the framework of scientific project No. \mbox {No.~19-32-90244}. The work on cataloging edge-on galaxies is carried out within the framework of RSF grant No.~19-12-00145.

\section*{CONFLICT OF INTEREST}
The authors declare no conflict of interest regarding the publication of this paper.

\appendix
%{APPENDIX}

\section{RFGC STRUCTURE} \label{a:RFGC}

The catalog is presented as a table, following the structure of
the RFGC catalog described in the original paper by \citet{RFGC}:
\begin{list}{}{
\setlength\leftmargin{4mm} \setlength\topsep{2mm}
\setlength\parsep{0mm} \setlength\itemsep{2mm} } \item {\it
rfgc}---the galaxy identifier from RFGC catalog in the range from
RFGC\,0001 to RFGC\,4236; \item {\it fgc}---the name of the galaxy
from the catalog FGC \citep{FGC} (FGC\,0001--FGC\,2573). The
southern extension of the catalog contains identifiers from
FGCE\,0001 to FGCE\,1882; \item {\it ra, dec}---the right
ascension and declination of the galaxy in degrees for the epoch
J2000.0. The coordinates have been refined with the HyperLeda
database \citep{HyperLeda}; \item {\it pa}--- the positional angle
of the major axis of the galaxy in degrees measured from north to
east; \item {\it aO, bO}---the ``blue'' diameters of the major and
minor axes of the galaxy in arc minutes in the POSS-I size system;
\item {\it aE, bE}---the ``red'' diameters of the major and minor
axes of the galaxy in arc minutes in the POSS-I size system; \item
{\it type}---the morphological galaxy type of a galaxy according
to the Hubble classification; \item {\it As}---the asymmetry index
(0---a symmetrical galaxy, 2---the pronounced asymmetry); \item
{\it sb}---the average surface brightness index (1---high,
4---very low); \item {\it Btot}---the total apparent $B$ magnitude
calculated based on the sizes, morphological type, and surface
brightness type as described in the RFGC catalog; \item {\it
nsat}---the number of significant satellites (see description in
the original RFGC catalog); \item {\it notes}---the notes on
galaxies; \item {\it pgc}---the PGC galaxy number from the
HyperLeda database \citep{HyperLeda}.
\end{list}

\section{EGIS STRUCTURE} \label{a:EGIS}

The catalog is generated based on the latest versions of
Tables\footnote{\url{http://users.apo.nmsu.edu/~dmbiz/EGIS/}} 4
and 6 from the paper by \citet{EGIS}. The original tables give
photometry of galaxies performed in two different ways. They have
been converted into three database tables to eliminate redundancy.

\subsection*{egis}

This table contains a list of ``true'' edge-on galaxies
\citep{EGIS}: \label{a:egis}
\begin{list}{}{
\setlength\leftmargin{4mm} \setlength\topsep{2mm}
\setlength\parsep{0mm} \setlength\itemsep{2mm} } \item {\it
eon}---the unique galaxy identifier from the original paper; \item
{\it ra, dec}---the right ascension and declination of a galaxy in
degrees for the epoch J2000.0; \item {\it altname}---the name of
the galaxy in other catalogs; \item {\it type}---the morphological
type of a galaxy: Sa, Sb, Sc, Sd, or Ir; \item {\it rv}---the
heliocentric radial velocity in km s$^{-1}$ according to HyperLeda
as listed in the original catalog \citep{EGIS}.
\end{list}

\subsection*{egis\_phot3d}

The photometry table from the one-dimensional analysis of the
brightness profiles of edge-on galaxies:
\begin{list}{}{
\setlength\leftmargin{4mm} \setlength\topsep{2mm}
\setlength\parsep{0mm} \setlength\itemsep{2mm} } \item {\it
eon}---the unique identifier associated with the \verb|egis|
Table; \item {\it band}---the SDSS filter($gri$); \item {\it
pa}---the position angle of the galaxy; \item {\it h, e\_h}~---the
radial exponential scale of the galaxy disk in arc seconds and its
error; \item {\it z0, e\_z0}---the vertical $\mathrm{sech}^2$
scale of the galaxy disk in arc seconds and its error; \item {\it
sb0, e\_sb0}---the central surface brightness of a galaxy modified
to the face-on position, in mag\,arcsec$^{-2}$ and its error;
\item {\it grad\_z0}---the gradient of the vertical scale of the
disk $z0$ normalized to the ratio of the scales:
$\dfrac{dz_0}{dr}\dfrac{h}{z_0}$; \item {\it mag}---the total
aperture magnitude inside the bounding ellipse corrected for
extinction in our Galaxy according to \citep{1998ApJ...500..525S};
\item {\it B/T}---the ratio of the bulge brightness to the total
luminosity of a galaxy; \item {\it fits}---the indicator to the
FITS file in the local archive; \item {\it ima}---the indicator to
the galaxy image in the given filter for visualization using the
web interface.
\end{list}

\subsection*{egis\_phot1d}

The photometry table from the analysis of the brightness
distribution of galaxies in the SDSS image in the $r$ filter
(3D-Analysis):
\begin{list}{}{
\setlength\leftmargin{4mm} \setlength\topsep{2mm}
\setlength\parsep{0mm} \setlength\itemsep{2mm} } \item {\it
eon}---the unique identifier associated with the \verb|egis|
Table; \item {\it band}---the SDSS filter. Should always be equal
to ($r$); \item {\it h}---the radial exponential scale of the
galaxy disk in arc seconds; \item {\it z0}---the vertical
$\mathrm{sech}^2$ scale of the galaxy disk in arc seconds; \item
{\it sb0}---the central surface brightness of the galaxy modified
to the face-on position, in mag\,arcsec$^2$.
\end{list}

\section{2MFGC STRUCTURE} \label{a:2MFGC}

\begin{list}{}{
\setlength\leftmargin{4mm} \setlength\topsep{2mm}
\setlength\parsep{0mm} \setlength\itemsep{2mm} } \item {\it
id}---the 2MFGC identifier of a galaxy \linebreak
(2MFGC\,00001--2MFGC\,18020); \item {\it pgc}---the PGC galaxy
number from the HyperLeda database \citep{HyperLeda}; \item {\it
ra, dec}---the right ascension and declination of a galaxy in
degrees for the epoch J2000.0; \item {\it r}---the elliptical Kron
radius in the 2MASS filter $K_s$. This aperture was used for
photometry in all three 2MASS filters.; \item {\it Jmag}---the
Kron magnitude in the \mbox {$J$ filter} of the 2MASS survey;
\item {\it Hmag}---the Kron magnitude in the \mbox {$H$ filter} of
the 2MASS survey; \item {\it Ksmag}---the Kron magnitude in the
\mbox {$K_s$ filter} of the 2MASS survey; \item {\it b/a}---the
axial ratio of a galaxy in the composite $J+H+K_s$ image; \item
{\it b/a1}---the axial ratio averaged over individual images in
the $J$-, $H$-, and $K_s$ filters; \item {\it pa}--- the
positional angle in the composite image measured from north N to
east E; \item {\it CI}---the concentration index in the $J$ filter
(the ratio of the radii in which 3/4 and 1/4 of the galaxy light
is concentrated).
\end{list}

\section{STRUCTURE OF TABLES OF CANDIDATES FROM THE PAN-STARRS SURVEY} \label{a:Pan-STARRS}

\subsection*{ps1candidate}

The table gives information about edge-on galaxy candidates found
in the Pan-STARRS survey. It contains the identifier of the
object; its coordinates; photometric parameters obtained during
primary selection, and indicators to files in the local archive.
\label{a:ps1candidate}
\begin{list}{}{
\setlength\leftmargin{4mm} \setlength\topsep{2mm}
\setlength\parsep{0mm} \setlength\itemsep{2mm} } \item {\it
projcell, subcell, candidate}---this three-digit combination is
used as the unique identifier of objects and is determined by the
specific character of the image archive organization in the
Pan-STARRS
survey\footnote{\url{https://outerspace.stsci.edu/display/PANSTARRS/PS1+Sky+tessellation+patterns}}.
The pair of numbers \verb|projcell| and \verb|subcell| indicates
the projection number in the sky and the cell number in the given
division, respectively. The candidate
number---\verb|candidate|,---found in this image by our search
algorithm for edge-on galaxies; \item {\it ra, dec}---the right
ascension and declination of the candidate in degrees for the
J2000.0 epoch; \item {\it sma\_r}---the major semi-axis
corresponding to the characteristic width of the Gaussian
inscribed in the two-dimensional distribution of light from a
galaxy in the $r$-filtered image; \item {\it ell\_r}---the
ellipticity, $1-b/a$, of the corresponding Gaussian; \item {\it
pa\_r}---the corresponding positional angle in the image; \item
{\it mag\_r}---the estimation of the total apparent magnitude of
the object; \item {\it fits\_g, fits\_r, fits\_i}---the indicators
to FITS files in the local image archive for each of the three
filters; \item {\it image, contour}---the color image and image
with the marked outline of the candidate selection, respectively,
for the convenience of viewing candidates.
\end{list}

\subsection*{ps1candidate\_annvote}

This table gives information on candidate classification obtained
in five different artificial neural network models trained to
classify edge-on galaxies:
\begin{list}{}{
\setlength\leftmargin{4mm} \setlength\topsep{2mm}
\setlength\parsep{0mm} \setlength\itemsep{2mm} } \item {\it
projcell, subcell, candidate}---the unique identifier of the
object associated with the \verb|ps1candidate| Table; \item {\it
n}---the number of a model used to classify a candidate; \item
{\it vote}---the index of conformity to edge-on galaxies obtained
within the framework of the corresponding model: 0---does not
match, 1---classified as an edge-on galaxy.
\end{list}

\subsection*{ps1candidate\_class}

This table gives information about the visual classification of
candidates by the project participants. The classification process
was divided into several stages, thus, the same object could be
classified by the same person several times.
\begin{list}{}{
\setlength\leftmargin{4mm} \setlength\topsep{2mm}
\setlength\parsep{0mm} \setlength\itemsep{2mm} } \item {\it
projcell, subcell, candidate}---the unique identifier of the
object associated with the \verb|ps1candidate| Table; \item {\it
userid}---the identifier of the user who performed the
classification; \item {\it date}---the classification time; \item
{\it workflow}---the indicator of the stage, during which the
identification was carried out; \item {\it class}---the actual
classification performed by this user during the current stage.
Possible values are: \verb|good|---a galaxy is almost edge-on;
\verb|acceptable|---a galaxy is seen at a high angle to the line
of sight; \verb|unsuitable|---an object is not an edge-on galaxy;
\verb|wrong|---a candidate is not a galaxy (a defect in the image
or a combination of stars); \item {\it use}---the flag of using
this classification in statistics (\verb|true| or \verb|false|).
\end{list}

\subsection*{ps1candidate\_phot}

The table presents the photometry of galaxies performed by the
SExtractor\footnote{\url{https://sextractor.readthedocs.io/}} code
with the images from the Pan-STARRS survey in five filters:
\begin{list}{}{
\setlength\leftmargin{4mm} \setlength\topsep{2mm}
\setlength\parsep{0mm} \setlength\itemsep{2mm} } \item {\it
projcell, subcell, candidate}---the unique identifier of an object
associated with the \verb|ps1candidate| Table; \item {\it
band}---the Pan-STARRS1 filter ($g$, $r$, $i$, $z$, and $y$);
\item {\it ra, dec}---the right ascension and declination of a
candidate in degrees for the epoch J2000.0; \item {\it xima,
yima}---the coordinates of the barycenter of an object in the
image (the parameters \verb|X_IMAGE|, \verb|Y_IMAGE| in
SExtractor); \item {\it aima, bima, paima}---the major and minor
semi-axes and the positional angle of the ellipse describing the
given object in the image (the parameters \verb|A_IMAGE|,
\verb|B_IMAGE|, \verb|THETA_IMAGE| in SExtractor); \item {\it a,
e\_a}---the semi-major axis of an object in the sky and its error
in arc seconds (the parameters \verb|A_IMAGE|, \verb|ERRA_IMAGE|);
\item {\it b, e\_b}---the minor semi-axis of an object in the sky
and its error in arc seconds (the parameters \verb|B_IMAGE|,
\verb|ERRB_IMAGE|); \item {\it pa}---the position angle of an
object measured from the north to east (the parameter
\verb|THETA_J2000|); \item {\it ell}---the ellipticity of the
object $1-b/a$ (the parameter \verb|ELLIPTICITY|); \item {\it
radkron}---the Kron pseudo-radius (the parameter
\verb|KRON_RADIUS|); \item {\it fluxauto, e\_fluxauto}--- the flux
after the background subtraction and its error inside the Kron
ellipse (the parameters \verb|FLUX_AUTO|, \verb|FLUXERR_AUTO|);
\item {\it magauto, e\_magauto}---the magnitude and its error
inside the Kron ellipse (the parameters \verb|MAG_AUTO|,
\verb|MAGERR_AUTO|); \item {\it radpetro}---the Petrosian
pseudo-radius (the parameter \verb|PETRO_RADIUS|); \item {\it
fluxpetro, e\_fluxpetro}---the flux after the background
subtraction and its error inside the the Petrosian ellipse (the
parameters \verb|FLUX_PETRO|, \verb|FLUXERR_PETRO|); \item {\it
magpetro, e\_magpetro}---the magnitude and its error inside the
Petrosian ellipse (the parameters \verb|MAG_PETRO|,
\verb|MAGERR_PETRO|); \item {\it badpixfraction}---the fraction of
``bad'' pixels inside the Petrosian ellipse describing a galaxy;
\item {\it quality}---the photometry quality obtained on the basis
of statistics of deviations of the ellipse parameters from the
median values; \item {\it fits}---the indicator of the file name
in the local archive.
\end{list}

\subsection*{ps1candidate\_crossid}

The table of cross-identification of candidates with galaxies from
the HyperLeda\footnote{\url{http://leda.univ-lyon1.fr/}} database
\citep{HyperLeda}:
\begin{list}{}{
\setlength\leftmargin{4mm} \setlength\topsep{2mm}
\setlength\parsep{0mm} \setlength\itemsep{2mm} } \item {\it
projcell, subcell, candidate}---the unique identifier of the
object associated with the \verb|ps1candidate| Table; \item {\it
ra, dec}---the right ascension and declination of the candidate in
degrees for the epoch J2000.0; \item {\it pgc}---the PGC galaxy
number from HyperLeda database \citep{HyperLeda}; \item {\it
flag}---the code of the identification type; \item {\it
userid}---the user who performed the identification; \item {\it
ctime}---the time of entering the identification into the
database.
\end{list}

\subsection*{ps1candidate\_notes} Various notes made during the work with candidates:
\begin{list}{}{
\setlength\leftmargin{4mm} \setlength\topsep{2mm}
\setlength\parsep{0mm} \setlength\itemsep{2mm} } \item {\it
projcell, subcell, candidate}---the unique identifier of an object
associated with the \verb|ps1candidate| Table; \item {\it
userid}---the user who made a note; \item {\it ctime}---the time
of making a note; \item {\it note}--- the note itself.
\end{list}

\section{STRUCTURE OF TABLES USED IN THE CLASSIFICATION SYSTEM} \label{a:VisualClassification}

\subsection*{user}

\begin{list}{}{
\setlength\leftmargin{4mm} \setlength\topsep{2mm}
\setlength\parsep{0mm} \setlength\itemsep{2mm} } \item {\it
id}---the unique login of a user; \item {\it name}---the full
username.
\end{list}

\subsection*{quiz}

\begin{list}{}{
\setlength\leftmargin{4mm} \setlength\topsep{2mm}
\setlength\parsep{0mm} \setlength\itemsep{2mm} } \item {\it
id}---the unique identifier of the questionnaire; \item {\it
title}---the short description; \item {\it tbl}---the name of the
table with objects to classify; \item {\it retired}---the maximum
number of classifications of an object by different users.
\end{list}

\subsection*{quiz\_question}

\begin{list}{}{
\setlength\leftmargin{4mm} \setlength\topsep{2mm}
\setlength\parsep{0mm} \setlength\itemsep{2mm} } \item {\it
id}---the unique number of a question; \item {\it quizid}---the
indicator of the questionnaire in the \verb|quiz| Table; \item
{\it value}---the assigned value of a characteristic; \item {\it
description}---the short description of a question; \item {\it
bunch}---the number of the question group; \item {\it
input}---determines the way of classification:
\verb|radio|---selection of one value from the set or
\verb|checkbox|---the presence or absence of this characteristic.
\end{list}

\onecolumngrid
\begin{flushright}
{\it Translated by N.~Oborina}
\end{flushright}
%\endinput

\end{document}